# Modeling temporal constraints for a system of interactive scores

Mauricio Toro*[1] , ºMyriam Desainte-Catherine, ºAntoine Allombert,

*Universidad EAFIT, Colombia

ºLaBRI, Université de Bordeaux, France

03 March 2017

**Abstract**

In this chapter we explain briefly the fundamentals of the interactive scores formalism. Then we develop a solution for implementing the ECO machine by mixing petri nets and constraints propagation. We also present another solution for implementing the ECO machine using concurrent constraint programming. Finally, we present an extension of interactive score with conditional branching.

## 1 Introduction

Stating as a definition of musical interpretation the possibility of interactively modifying some parameters of a piece of music during its execution, leads to technical and theoretical problems in the context of electroacoustic music. Indeed this type of music is made withany sort of sounds, and the composition process consists of collecting and producing a sound material before temporally organizing it. Therefore, a piece of electroacoustic music is fixed on a physical medium and broadcasted during concerts. Consequently, a musician cannot give his own interpretation of a piece such as he could do so with a traditional piece of music.

In [ALL 09] we explored the opportunity of adapting the traditional way to interpret music to electroacoustic music. We limited ourselves to the question of *agogic* modifications (i.e. the possibility, during an execution, to modify the starting or ending moments of the notes of a traditional score). We proposed a formalism of composition with which a composer can express some liberties and limits to *agogic* modifications, by using temporal constraints. We call *interactive scores*, the score composed using this formalism. We also presented solutions for implementing an abstract machine (the *ECO machine*), able to execute *interactive scores*, that respects the constraints given by the composer and allows the musician to take benefits from the interpretation liberties.

In this chapter we explain briefly the fundamentals of the *interactive scores* formalism in section 1.2. Then we develop a solution for implementing the *ECO machine* by mixing petri nets and constraints propagation in section 1.3. In section 1.4, we present another solution for implementing the *ECO machine* using concurrent constraint programming. We present an extension of *interactive score* with conditional branching in section 1.5. Finally, we present some concluding remarks and future work in section 1.6.

---

1 Corresponding author. Email: mtoro [at] eafit.edu.co

## 2 Formalism of *interactive scores*

We fully presented the *interactive score* formalism in [ALL 09]. Let us remember some definitions on which this chapter is based.

This formalism stems from an analyze of the interpretation of traditional music, that enlightened that a composer can specify through the score qualitative and quantitative temporal constraints between the starting and ending dates of notes. A composer can also specify some possibilities of modifications. The formalism is hierarchical. Therefore, a basic element of interactive score, the *temporal objects* can be either complex (it carries other *temporal objects*), in this case it is called a *structure*, or simple and called a *texture*.

Formally, a *temporal object* is defined as a 11-uple :

$$TO=<t,r,p,E,S,Pc,V,En,B,R,C>$$

We shall only develop the useful points for this chapter. Please refer to the previously cited document for further explanations. Figure 2.3.3 presents an example of an *interactive score*. A structure is associated to a time referential which is left to right oriented.

- *p* : a process associated to *TO*, only if *TO* is a *texture*, null otherwise.

- *Pc* : a set of dated *control points* ; the *control points* represent particular moment of the execution of *TO*, on the figure 2.3.3, they are represented by big circles on the up and down borders of the objects ; if *TO* is a *texture*, its *control points* represent computation steps of its process *p* (its starting or ending date, or for example a particular value of a table ; by giving a date to each *control point* of a *texture*, the user can specify the temporal proprieties of the execution of the process *p* ; if *TO* is a *texture* a control point represents a *control point* of one of the *TO*'s children, in the purpose of defining temporal relations with objects in an upper level ; the dates of the control points are expressed in the time unit of parent *structure* of *TO*.

- *En* : if *TO* is a *structure*, the set of the children objects of *TO*, null otherwise.

### 2.1 Temporal relations

A composer can define the temporal proprieties of a score through temporal relations between the *control points* of the *temporal objects*. Thus these relations are taken from the point algebra. There are 2 possible relations : *precedence* (*Pre*) and *posteriority* (*Post*). *Pre* and *Post* are symmetrical and constitute temporal qualitative constraints over the *control points*. In addition, the user can specify quantitative constraints, by giving a range of possible values for the time interval between two points bound by a relation.

A relation *tr* of the set *R* of a *structure S* is defined by a 5-uple :

- *t* is a type (*Pre* or *Post*)
-  and  are *events* of *S*
-  and  are real values in [0,∞]

If *tr* is a precedence relation, then it imposes the inequality :

where is the date of .

Remember that these dates are expressed in a the time referential *S*, so are expressed the values and .

## 2.2 Interaction Points

A composer can associate an *interaction point* (graphically denoted by *Pi*) to a *control point*, in order to define the *control point* to be dynamically triggered by a musician during execution. This type of *control points* are said *dynamic* (as the beginning of *texture* A in the *structure* of figure 2.3.3), while the other *control points*, the *static* ones, will be triggered by the system. The written date of a dynamic *control point* is clearly indicative, since the performer can modify it. However, this modification possibility is limited the temporal relations. Then, on figure 2.3.3, since there is an implicit *precedence* relation between and , the system will refuse the dynamic triggering of before . In like manner, the system will automatically trigger if the musician wait so long that some minimum values of intervals cannot be respected. We will develop this type of situations in the next section.

## 2.3 Modification behaviors

During execution of a score, modifications of the written date of a dynamic *control point* can lead to modifications of the value of some time intervals, in order to respect the temporal constraints defined by the composer. As an example, on figure 2.3.3, to delay increases the duration of the time interval between and . This increase must be spread to the other durations. For example, we could increase in the same proportion (the maximum value of will limit the delay on ). Or we could dwindle the duration of some time intervals inside *S*. Or we could do both. We propose four strategies for spreading a modification of a duration to other time interval durations. We call them *modification behaviors*. In order to present these behaviors, let us introduce two notations on the example of figure 2.3.3. As mentioned before, there are some implicit temporal relations between a *structure* and its children. There is a *precedence* relation between the starting date of a *structure* and the starting date of each child, and there is a *precedence* relation between the end of each child and the end the *structure*. Then, we will denote by the duration between and , and the duration between and . In addition, we suppose that both of these durations are *supple* (i.e. their minimum value is 0 and their maximum value is +∞). Note that a composer could have limit them with quantitative constraints.

The strategy of each behavior is based on an equality between the duration of an interval and the sum of the duration of its sub-intervals. As an example, for the score of figure 2.3.3, this equality is the following one :

$$\tag{1}$$

Each behavior refers to this equality to spread a modification of a duration to other durations.

### 2.3.1 *Fermata* behavior

In the traditional music notation, a *fermata* is a sign associated to a note, that allows a musician to make last the note "as long as he wants". When a musician takes benefit of such a liberty, the notes following the modified note will be temporally shifted.

They will be shifted backward if the note is shortened, and forward if the note is prolonged. We directly adapt the *fermata* notation into a so called behavior. According to this behavior, if the date of a *control point* is modified, the following *control points* and *temporal objects* are shifted, and the duration of the parent *structure* that includes the modified *control point* is accordingly modified. Taking figure 2.3.3 as a score, one can find on figure the consequence of a delay on with a *fermata* behavior. is increased and objects *A* and *B* are moved back without any modification of , , or , while is increased according to the equality 1. The value will limit the possibility of increasing and therefore the possibility of increasing . During the execution, the system will automatically trigger if the musician wait so long that reaches . In like manner, the system will prevent the musician from triggering so early, that the dwindling of leads to be lesser that .

### 2.3.2 *Chronological* and *anti-chronological* behaviors

The *chronological* and *anti-chronological* behaviors are two strategies that try to spread a modification of a duration to each duration that can be modified. They differ on the order in which they modify the durations. The *chronological* behavior will spread a modification starting by the right-hand part of the equality 1 (i.e. the sub-intervals), following the chronological order (the order in which the sub-intervals appear in equality 1). The left-hand part of the equality is modified last. Figures Error: Reference source not found and Error: Reference source not found present the first two step of an increase of . First, is dwindled to its minimum, secondly is dwindled to its minimum. If is still delayed, the system will dwindle and . At last, the system will increase to accept the increase of . This will lead to the situation presented on figure Error: Reference source not found. On the contrary, the *anti-chronological* behavior starts by modifying the left-hand part of the equality 1 (), before modifying the right-hand part in an anti-chronological order (i.e. the reverse order in which the durations appear in equality 1). Figures Error: Reference source not found and Error: Reference source not found present the first two steps of the spread of an increase of . First, is increased, secondly is dwindled to its minimum (which is 0). If is still delayed, and will be dwindled. A last, this behavior will lead to the same situation than the *chronological* one (the situation presented on figure Error: Reference source not found). For both behaviors, the maximum and minimum values of the durations will limit the modification possibilities. Thus, the system will trigger if it reaches the final situation (nothing more can be modified). The system will also prevent from triggering too early.

### 2.3.3 Proportional behavior

The proportional behavior can be chosen only if the left-hand part of the equality 1 id *rigid* (i.e. its minimum and maximum value are the same, so it cannot be modified). This behavior only modifies the right-hand part of the equality, such that each duration is modified in the same proportion. Figure presents the spread of an increase of .

During the composition process, the composer can choose for each *structure*, the behavior that will be used to spread during the execution a modification of a duration of the *structure*.

# 3  ECO Machine

The *ECO* machine is the abstract machine that executes the *interactive score*. During an execution of a score this machine maintains the constraints defined by the composer, it allows the musician to use the interaction points to interpret the score, and it spreads modifications of durations according to the chosen behaviors. We present possible implementations of the *ECO* machine based on petri nets. The first one corresponds to a specific case in which we reduce the temporal constraints and behaviors that a composer can define. The second implementation is an extension of the first one to accept all kinds of constraints and behaviors.

## 3.1  *Supple* intervals with *fermata* behavior

For the first implementation, we impose that each interval must be *supple* (i.e. their minimum value is 0 and their maximum value is +∞), and that the modification behavior of each *structure* is the *fermata* behavior. In this case, it is possible to translate an *interactive score* into a petri net. We use *hierarchical time flush* petri nets . Each *structure* is turn into a petri net, the set of petri nets produced by this operation, is organized according to the same hierarchical organization than the *interactive score* to create the *hierarchical* petri net. Translating a *structure* consists of turning each *control point* into a transition, and each interval into a place between the transitions representing the *control points* that the interval separates. Figure 3.1 presents an example of *structure* and its translation into a petri net.

Executing the interactive score consists of running the petri net. When a transition is crossed, the process step associated to the *control point* represented by the transition is executed. This produces data that are routed to other applications in order to broadcast the content of the score. Crossing a transition that represents a dynamic *control point*, depends on a triggering action of the musician. Since there are no quantitative constraints in this case, the qualitative constraints are maintained by running rules of the petri net. Indeed before crossing a transition $t$, all transitions that precede $t$ must have been crossed. Concerning the static control points, the arcs of the *time flush* petri nets that income into a transition, can be labeled with a range of values and a nominal value. On the example of figure 3.1, let us denote by $P$ the place between transition  and transition . If the arc between $P$ and  is labeled with a range of values  and a nominal value , then by denoting the crossing date of a transition $t$ by $d(t)$ , the execution rules of the petri net impose :

 is the ideal value of this difference of dates. During the translation of a score into a petri net, an arc outgoing from a place $P$ is labeled with the range [0,+∞]. The nominal value of this arc is the written value of this interval represented by $P$. With this type of labels, the execution of the petri net complies with the temporal constraints of the score and the *fermata* behavior.

## 3.2  General case

Taking into account the different modification behaviors and the quantitative constraints on the durations, leads to extend the previous solution. Even if it is possible to make a petri net modifying the labels of its arcs, encoding the strategies of the behaviors directly into a petri net seems to be impossible. We tried this way using colored petri nets, but this solution appeared to be a deadlock.

We propose a general method which consists of adding a constraints solver system to the petri net.

Two questions arise :

- which type of constraints must we use ?

- which type of methods must we use to solve the constraints problem ?

To explain the answers to these questions, we will lean on the example of figure 3.1. To avoid a long definition of all the intervals of this example, we directly present on figure 3.2 the graph of events of this score. This type of graph represents events (in our the control points) as nodes, and time intervals between the events as labeled arcs. We can see on figure 3.2 that all explicit and implicit relations appear. We also suppose that the composer has defined a quantitative constraints on  and , that's why they appear as arcs in the graph.

## 3.3 Interval constraints

The definition of the modification behaviors was caused by the observation that the composer can specify quantitative constraints on intervals that can be divided in several sub-intervals. In the example of figure 2.3.3,  is considered as a constrained interval, while the intervals inside *S* are considered as the sub-intervals of . According to the equality 1, a modification of one of the intervals appearing in this equality must be spread to the rest of the intervals. Thus, it is clear that we must use constraints of the form of equality 1. We call this type of constraints *interval constraints*. There is another case to take into account, the equality of two right-hand parts of equality without an explicit definition of the left-hand part. As an example, considering on the events graphs of figure 3.1 the duration between  and  we can deduce the equality :

(2)

by introducing an explicit variable , we can turn equality 2 into two equalities of the type of 1. This type of constraints must be taken into account if one of the interval of equality 2 is not *supple*, because in this case, the introduced variable  will not be *supple*. If the left-hand part of an equality of the type of equality 1 is *supple*, the petri net can manage the constraints.

For each structure *S* of an *interactive score*, we want to build a constraint graph only made of constraints of the type equality 1. We must identify in *S* all relevant equalities of type 1 and 2, in the second case we introduce a variable and produce two constraints of type 1. Our technique is based on the events graph of *S*, . Due to the implicit temporal relation in *S*,  has necessarily two terminals ( and ). In addition, one can see that equalities of type 1 and 2 can be identified when two branches of  start with the same vertex, and end with the same vertex. Thus we can deduce that these equalities can be identified in the graph as some *series-parallel* sub-graphs of . Then, to identify the relevant equalities, we build a *series-parallel* sub-graph Γ of  that includes  and , and that maximizes its number of vertices. Since  and  are terminals in , Γ necessarily exists, but is not necessarily unique. Figure 3.3 gives such a sub-graph for the graph of figure 3.2.

It is clear how to deduce equalities of form 1 and 2 from this sub-graph. Naturally, is not necessarily a series-parallel graph and is not necessarily equal to . We can take into account the equalities missing in by considering the branches of that are not in , and by adding constraints for each of them. Figure presents a constraints graph associated to the structure of figure 3.1 (for the moment please do not take the orientation into account). is based on the sub-graph of figure 3.3. One can see that the constraints graph that only contains the equalities of do not have cycles. The equalities added by the branches of that are not in , introduce cycles in . Thus, the bigger is , the lesser is the number of cycles in .

## 3.4 Constraints propagation

We choose to use a propagation algorithm to solve the constraints problem during the execution of a score. This choice is based on several observations : the values of durations written in the score constitute a solution to the constraints problem ; triggering a dynamic control point introduces a perturbation in the solution of the written score ; for a constraint $c$, the choice of the propagation method depends on the variable of $c$ which is modified ; there is no addition of constraints during the execution of a score ; there is no addition of control points during the execution of a score.

As a consequence, it is possible to orient the constraints graph for each dynamic control point, in order to represent how the modification of a dynamic control point will be spread through the other intervals. Such an orientation is presented on figure . This orientation corresponds to the modification of by the triggering . One must notice that the methods of propagation must take into account the execution time. This means during the propagation in a constraint $c$, the value of some variables of $c$ can be fixed, because the control points that delimit the intervals represented by the variables occurred. Once this has been taken into account, it is possible to use a *projection based compilation*. This consists of statically analyzing each perturbation that can occur during thanks to the constraints graph, and in producing a list of actions for spreading the perturbation when it occurs. During the execution, when a dynamic control point is triggered, we launch the actions to spread the modification.

Finally the general case solution implementation can be sum up as following :

- translating the score into a petri net with the rules than the *fermata* behaviors case.

- for each structure $S$ , building the events graph of $S$ (one can see that this graph is very close from the petri net built for $S$).

- for each structure $S$, using to build the constraints graph of $S$ by finding a series-parallel sub-graph of .

- for each structure $S$, for each dynamic control point $P$ of $S$, using to produce a list of actions that will spread the modification introduced by the triggering of $P$.

During the execution of the score, the petri net is run. When a dynamic control point is triggered, its date of triggering is communicated to the propagation system. This system launches the actions that spread the modifications. The petri net is modified with the new values of intervals computed by the propagation system.

# 4 Concurrent Constraints Time Model

In this section we present another model of interactive scores, based upon *Non-deterministic Timed Concurrent Constraint* (`ntcc`) [NIE 02] calculus. Process calculi has been applied to the modeling of interactive music systems [TOR 16b, TOR 16d, Tor 16a, TOR 15c, ALL 11, TOR 14a, TOR 09c, OLA 11, TOR 12a, TOR 09a, TOR 10a, TOR 15a, ARA 09, TOR 12b, TOR 09b, TOR 10b, TOR 10c, TOR 08] and ecological systems [TOR 16c, PHI 13a, TOR 14b, PHI 13b, TOR 15b]. An advantage of `ntcc` is that process synchronization is achieved by adding or deducing constraints from a constraint store, thus is declarative.

The model is inspired on a previous model that does not consider hierarchy nor the fact that execution can continue even if interactive events are not launched [ALL 06]. The model is close in spirit to the *petri nets model*. Regrettably, in petri nets it is difficult to model global constraints such as the maximum number of simultaneous temporal objects and temporal reductions during execution.

Another related model is *Tempo*, a formalism to define declaratively partial orders of musical and audio processes [RAM 06]. Unfortunately, Tempo does not allow us to express choice when multiple conditions hold, simultaneity nor to perform an action if a condition cannot be deduced.

## 4.1 Ntcc model

We use `ntcc` to express operational semantics of interactive scores. In order to define operational semantics, we need to transform an interactive score into a graph where the vertices are the starting and ending points of temporal objects and arcs represent the delays among them, whereas in petri nets this is explicit. We define the graph as $g=(V,A,lV,lA)$ where $V$ is the set of vertices, $A$ the set of arcs, $lV$ a function that assigns labels to vertices, $lA$ a function that assigns labels to arcs. A vertex is labeled with the type of control point (a starting point, ending point or interactive point) and the temporal object associated to it. An arc is labeled by its duration. The graph is reduced by removing zero-delay arcs and representing several points in the same vertex.

Absence of zero delays simplifies the definition of operational semantics because we do not have to synchronize two processes to occur at the same time. To remove a zero-delay arc between a vertex $a$ and $b$, we delete $a$ and we connect all its successors and predecessors to $b$. We also combine the label of $a$ with the label of $b$, this means that a vertex may represent the starting or ending of several points.

### 4.1.1 Control Points

In this model we have two type of control points: interactive (*iPoint*) and static points (*sPoint*). Process updates the variable with `true` and persistently assigns to the current value of *clock*. Event is the user event associated to point $i$: it represents a user interaction. Set *Pr* represents the predecessors of $i$. Process *User* persistently chooses between launching or not an interactive event.

    when do (
 when do next
|unless next when do )
|unless next
    when do (

**unless next**
**|when do next** )
**|unless next**

**!(tell** () + **skip)**

### 4.1.2 Temporal relations

An relation between points *i* and *j* with a duration of ∆ is represented by the constraint
. The value of ∞ is approximated by the a parameter of the model, namely .
**!tell**

### 4.1.3 Example

Figure 2 represents the graph associated to the score in Figure 1. It does not contain zero-delays.

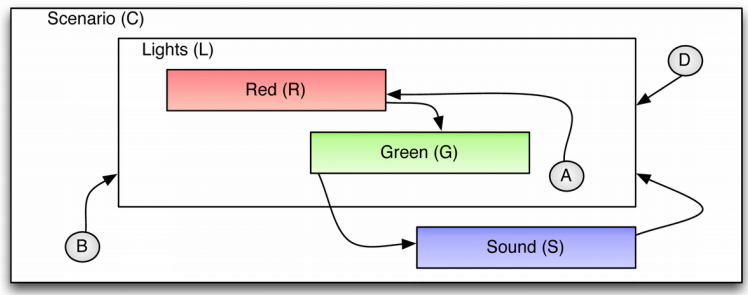

0.55

Figure 1: Example of a score.

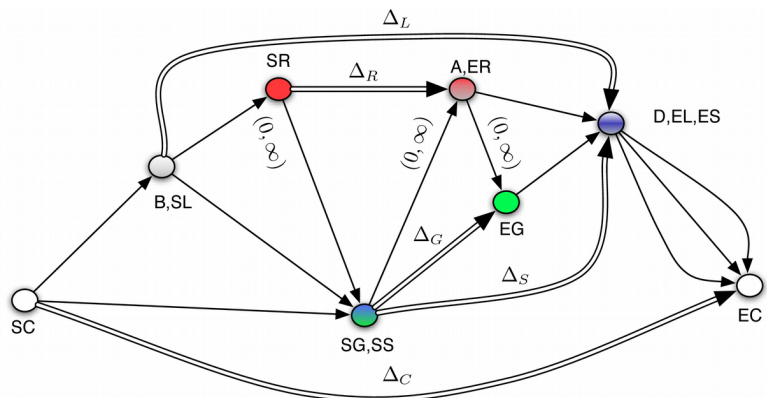

0.55

Figure 2: Graph representing the score in Fig. 1. Label denotes the starting point of *t* and the ending point of *t*. Double arrows represent the duration of temporal objects. Simple arrows with no label represent the constraints imposed by the hierarchy, its duration is [0,∞).1

CLOCK is a process that *ticks* forever. Process *Score* is parametrized by the graph in Fig. 2. We post a constraint to allow that be eventually deduced even in absence of user interactions.

*CLOCK*(*k*) **tell** (*clock=k*) | **next** *CLOCK*(*k*+1) |

## 4.2 Discussion

The interactive scores model coded in `ntcc` is not yet implemented. We plan to implement our system using *Ntccrt* [TOR 09c], a real-time capable interpreter for `ntcc`. There is an issue with the correctness of the implementation: it heavily relies on propagation. Each time an interactive point is launched, we add a constraint, and we propagate. Does propagation preserves the coherence of the model? Do we need to perform search or is propagation enough? This is an open issue. We also believe that the denotation of a score will help us to understand the behavior of the score without analyzing its operational semantics, to specify properties, and to prove its correctness.

# 5 Timed Conditional Branching Model

There is neither a formal model nor a special-purpose application to support conditional branching in interactive music. Using conditional branching, a designer can create pieces with choices, such as pieces in *open form*. The musician and the system can take decisions during performance with the degree of freedom described by the designer, and define when a loop ends; for instance, when the musician changes the value of a variable.

In the domain of composition of interactive music, there are applications such as *Ableton Live*[2]. Using *Live*, a composer can write loops and a musician can control different parameters of the piece during performance. Unfortunately, the means of interaction and the synchronization patterns are limited.

To express complex synchronization patterns and conditions, it was shown in [TOR 10b] that interactive scores can describe temporal relations together with conditional branching. In this section, we recall the conditional-branching timed model of interactive scores based upon the *Non-deterministic Timed Concurrent Constraint* (`ntcc`) calculus. We explain how to model temporal relations, conditional branching and discrete interactive events in a single model. We also present performance results of a prototype and we discuss the advantages and limitations of the model.

## 5.1 Specification of the model

A *score* is defined by a tuple composed by a set of points and a set of intervals. In this model, temporal relations are replaced by intervals that model both temporal relations and conditional branching. A temporal object is a type of interval.

2 http://www.ableton.com/live/

### 5.1.1 Control Points

We say that a *control point p* is a *predecessor* of *q* if there is a relation *p before q*. Analogically, a point *p* is a *successor* of *r* if there is a relation *r before p*. A Point is defined by , where  and  represent the behavior of the point. Behavior  defines whether the point *waits until all* its predecessors transfer the control to it or it only *wait for the first* of them. Behavior  defines whether the point *chooses* one successor which condition holds to transfer the control to it, or it does *not choose*, transferring the control to all its successors.

### 5.1.2 Intervals

An *interval p before q* means that the system waits a certain time to transfer the control (*jumps*) from *p* to *q* if the condition in the interval holds, and it executes a process throughout its duration. An interval is composed by a starting point, an ending point, a condition, a duration, an interpretation for the condition, a local constraint, a process, parameters for the process, children, and local variables.

There are two types of intervals. *Timed conditional relations* have a condition and an interpretation, but they do not have children, their local constraint is `true`, and their process is *silence*[3]. *Temporal objects* may have children, local variables and a local constraint, but their condition is `true`, and their interpretation is "*when* the condition is true, transfer the control from the starting point to the ending point".

A *timed conditional relation* includes the starting and ending points involved in the relation and the condition that determines whether the control *jumps* from starting to the ending point. There are two possible values for the interpretation of the condition: (i) *when* means that if condition holds, the control jumps to the end point; and (ii) *unless* means that if the condition does not hold or its value cannot be deduced from the environment, the control jumps to end point. A *temporal object* is an interval where the starting point launches a new instance of a temporal object and the ending point finishes such instance. Its local variables and local constraint can be used by its children and process to synchronize each other.

### 5.1.3 Example

Figure 3 describes a score with a loop. During the execution, the system plays a silence of one second. After the silence, it plays the sound *B* during three seconds and simultaneously it turns on the lights *D* for one second. After the sound *B*, it plays a silence of one second, then it plays video *B*. If the variable *finish* becomes true, it ends the scenario after playing the video *C*; otherwise, it jumps back to the beginning of the first silence after playing the video *C*.

---

3 *silence* is a process that does nothing.

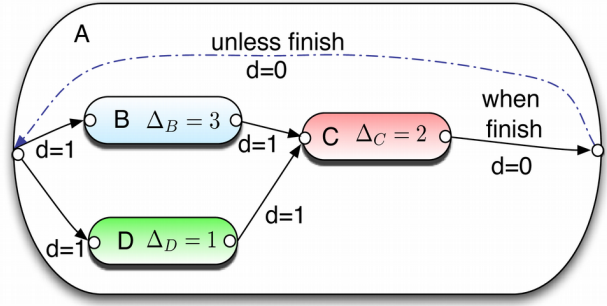

Figure 3: A score with a user-controlled loop.

The points have the following behavior. The end point of $C$ is enabled for choice, and the other points transfer the control to all their successors. The starting point of $C$ waits for all its predecessors and all the other points only wait for the first predecessor that transfers the control to them. Object $A$'s process is *silence*, its children are $B$, $C$ and $D$ and its local variable is *finish*. Note that the silence between $D$ and $C$ lasts longer during execution because of the behavior of the points.

## 5.2 Ntcc model

We model points and intervals as processes in `ntcc`. The definition of an interval can be used for both timed conditional relations and temporal objects. We represent the score as a graph; however, there is not a procedure to remove zero-delay arcs. Predecessors and successors of point $i$ are  and , respectively. For simplicity, we do not include hierarchy, we only model the interpretation *when*, we can only declare a single interval between two points, and we can only execute a single instance of an interval at the same time.

### 5.2.1 Control Points

We only model points that choose among their successors (*ChoicePoint*), points that transfer the control to all their successors (*JumpToAllPoint*), and points that wait for all their predecessors to transfer the control to them (*WaitForAllPoint*).

To know if at least one point has transferred the control to the point $i$, we ask to the store if  can be deduced from the store. When all the expected predecessors transfer the control to the point $i$, we add a constraint . Analogaly, when a point $i$ transfers the control to a point $j$, we add the constraint *ControlTranferred*($j$,$i$). In order to represent choice in the example of Fig. 1, we use the constraint *finish*.
 **! when do** (**tell**
| **when** *finish* **do tell** (*ControlTransferred*(*a*,*i*))
+**when** ≠*finish* **do tell** (*ControlTransferred*(*b*,*i*))) The following definition uses the parallel composition agent  to transfer the control to all the successors of the point $i$.
**when** *Succs*(*i*,*j*) **do tell** (*ControlTransferred*(*j*,*i*)))
 |**tell** Using the definition , we define the two types of point that transfer the control to all its successors. To wait for all the predecessors, we ask the store if the constraint *Arrived=Predec* holds.   **! when do**

### 5.2.2 Intervals

Process *I* waits until at least one point transfers the control to its starting point *i*, and at least one point has chosen to transfer the control to *j*. Afterwards, it waits until the duration of the interval is over[4]. Finally, it transfers the control from *i* to *j*.   !(**tell** (*Predec*(*j*,*i*)) | **tell** (*Succ*(*i*,*j*)))
 |**! when do**(
**next**(**tell**(*Arrived*(*j*,*i*)) |*PredecessorsWait*(*i*,*j*)))
    *PredecessorsWait* adds the constraint *Arrived*(*j*,*i*) until the point *j* is active.

### 5.2.3 Example

We can represent the score in Fig. 3 in `ntcc`. The starting point of *C* waits for all the points, the end of *C* chooses a point and the other points jump to all points. The intervals have the duration described in Fig. 3. We also need to model a user making choices. *User* tells to the store that *finish* is not true during the first *n* time units, then it tells that *finish* is true. It is initialized with *i*=0. An advantage of `ntcc` is that the constraint $i \geq n$ can be easily replaced by more complex ones; for instance, "there are only three active points at this moment in the score".
  **when** $i \geq n$ **do tell** (*finish*) |**unless** $i \geq n$ **next tell** ($\neq$*finish*)
 |**next**

## 5.3 Results and Discussion

Performance results are described in [TOR 10b]. They ran a prototype of the model over *Ntccrt*. The tests were performed on an iMac 2.6 GHz with 2 GB of RAM under Mac OS 10.5.7. They compiled it with GCC 4.2 and Gecode 3.2.2. The authors of the Continuator [PAC 02] argue that a multimedia interaction system with a response time less than 30 ms is able to interact in real-time with even a very fast guitar player. Response time was less than 30ms for the conditional branching model for up to 500 temporal objects.

## 6 Concluding Remarks and Future Work

We presented a model for interactive scores in petri nets. In order to represent temporal reductions, we have a constraint graph that gives information to the petri net to decide the duration of certain intervals. To our knowledge this is the first model that combines petri nets and temporal constraint programming in an interactive setting, and we think that it is worth to study this combination for many other problems involving temporal constraints.

We also presented a model in `ntcc` to represent temporal reductions and interactive events. The model is close in spirit to the petri nets model. An advantage of petri nets is that transformation from interactive scores to petri nets is simpler, and synchronization is easier when it depends, for instance, on the transitions that precede a place. An advantage of `ntcc` is that it can easily represent global constraints such number of simultaneous temporal constraints and also temporal reductions.

Finally, we presented a model of interactive scores with conditional branching in `ntcc` . An advantage of `ntcc` with respect to other models of interactive scores, Pure Data (PD), Max and Petri Nets is representing declaratively conditions by the means of constraints. Complex conditions, in particular those with an unknown number of parameters, are difficult to model in Max or PD [PUC 98] . To model generic conditions in Max or PD, we would have to define each condition either in a new patch or in a predefined library. In Petri nets, we would have to define a net for each condition. A disadvantage of this model is that we cannot always synchronize two objects to happen at the same time: one reason is choice and the other is that using *jumps* is not always possible to respect the constraints on their durations.

---

4 *next* is a process *next* nested *d* times (*next*(*next*(*next*...).

In the future we want to include sound synthesis and stream processing into the formalism of interactive scores. We want to explore other possible representations for interactive scores such as *synchronous languages*. Finally, we also want to develop automatic verification tools.

# References

[ALL 06]     ALLOMBERT A., ASSAYAG G., DESAINTE-CATHERINE M.RUEDA C., Concurrent Constraint Models for Interactive Scores, *Proc. of SMC '06*, May 2006.

[ALL 09]     ALLOMBERT A., Aspects temporels d'un système de partitions musicales interactives pour la composition et l'interprétation, PhD thesis, Université de Bordeaux, 2009.

[ALL 11]     ALLOMBERT A., DESAINTE-CATHERINE M.TORO M., Modeling Temporal Constrains for a System of Interactive Score, ASSAYAG G.TRUCHET C., , *Constraint Programming in Music*, 1, 1–23, Wiley, 2011.

[ARA 09]     ARANDA J., ASSAYAG G., OLARTE C., PÉREZ J. A., RUEDA C., TORO M.VALENCIA F. D., An Overview of FORCES: An INRIA Project on Declarative Formalisms for Emergent Systems, HILL P. M.WARREN D. S., , *Logic Programming, 25th International Conference, ICLP 2009, Pasadena, CA, USA, July 14-17, 2009. Proceedings*, 5649 *Lecture Notes in Computer Science*, Springer, 509–513, 2009.

[NIE 02]NIELSEN M., PALAMIDESSI C.VALENCIA F., Temporal Concurrent Constraint Programming: Denotation, Logic and Applications, *Nordic Journal of Comp.*, 1, 2002.

[OLA 11]     OLARTE C., RUEDA C., SARRIA G., TORO M.VALENCIA F., Concurrent Constraints Models of Music Interaction, ASSAYAG G.TRUCHET C., , *Constraint Programming in Music*, 6, 133–153, Wiley, Hoboken, NJ, USA., 2011.

[PAC 02]     PACHET F., Playing with Virtual Musicians: the Continuator in Practice, *IEEE Multimedia*, 9, 77–82, 2002.

[PHI 13a]    PHILIPPOU A.TORO M., Process Ordering in a Process Calculus for Spatially-Explicit Ecological Models., *Proceedings of MOKMASD'13*, LNCS 8368, Springer, 345–361, 2013.

[PHI 13b]    PHILIPPOU A., TORO M.ANTONAKI M., Simulation and Verification for a Process Calculus for Spatially-Explicit Ecological Models, *Scientific Annals of Computer Science*, 23, 1, 119–167, 2013.

[PUC 98]     PUCKETTE M., APEL T.ZICARELLI D., Real-time audio analysis tools for Pd and MSP, *Proc. of ICMC '98*, 1998.

[RAM 06]     RAMIREZ R., A Logic-based Language for Modeling and Verifying Musical Processes, *Proc. of ICMC '06*, 2006.

[TOR 08]     TORO M., Exploring the possibilities and limitations of Concurrent Programming for Multimedia Interaction and graphical representations to solve musical CSP's,  2008-3, Ircam, Paris.(FRANCE), 2008.

[TOR 09a]   TORO M., Probabilistic Extension to the Factor Oracle Model for Music Improvisation, Master's thesis, Pontificia Universidad Javeriana Cali, Colombia, 2009.

[TOR 09b]   TORO M., Towards a correct and efficient implementation of simulation and verification tools for Probabilistic ntcc, , Pontificia Universidad Javeriana, May 2009.

[TOR 09c]   TORO M., AGÓN C., ASSAYAG G.RUEDA C., Ntccrt: A concurrent constraint framework for real-time interaction, *Proc. of ICMC '09*, Montreal, Canada, 2009.

[TOR 10a]   TORO M., Structured Interactive Musical Scores, HERMENEGILDO M. V.SCHAUB T., , *Technical Communications of the 26th International Conference on Logic Programming, ICLP 2010, July 16-19, 2010, Edinburgh, Scotland, UK*, 7 *LIPIcs*, Schloss Dagstuhl - Leibniz-Zentrum fuer Informatik, 300–302, 2010.

[TOR 10b]   TORO M.DESAINTE-CATHERINE M., Concurrent Constraint Conditional Branching Interactive Scores, *Proc. of SMC '10*, Barcelona, Spain, 2010.

[TOR 10c]   TORO M., DESAINTE-CATHERINE M.BALTAZAR P., A Model for Interactive Scores with Temporal Constraints and Conditional Branching, *Proc. of Journées d'Informatique Musical (JIM) '10*, May 2010.

[TOR 12a]   TORO M., Structured Interactive Scores: From a simple structural description of a multimedia scenario to a real-time capable implementation with formal semantics , PhD thesis, Univeristé de Bordeaux 1, France, 2012.

[TOR 12b]   TORO M., DESAINTE-CATHERINE M.CASTET J., An Extension of Interactive Scores for Multimedia Scenarios with Temporal Relations for Micro and Macro Controls, *Proc. of Sound and Music Computing (SMC) '12*, Copenhagen, Denmark, July 2012.

[TOR 14a]   TORO M., DESAINTE-CATHERINE M.RUEDA C., Formal semantics for interactive music scores: a framework to design, specify properties and execute interactive scenarios, *Journal of Mathematics and Music*, 8, 1, 93–112, 2014.

[TOR 14b]   TORO M., PHILIPPOU A., KASSARA C.SFENTHOURAKIS S., Synchronous Parallel Composition in a Process Calculus for Ecological Models, CIOBANU G.MÉRY D., , *Proceedings of the 11th International Colloquium on Theoretical Aspects of Computing - ICTAC 2014, Bucharest, Romania, September 17-19*, 8687 *Lecture Notes in Computer Science*, Springer, 424–441, 2014.

[TOR 15a]   TORO M., Structured Interactive Music Scores, *CoRR*, abs/1508.05559, 2015.

[TOR 15b]   TORO M., PHILIPPOU A., ARBOLEDA S., VÉLEZ C.PUERTA M., Mean-field semantics for a Process Calculus for Spatially-Explicit Ecological Models, , Department of Informatics and Systems, Universidad Eafit, 2015, Available at `http://blogs.eafit.edu.co/giditic-software/2015/10/01/mean-field/`.

[TOR 15c]   TORO M., RUEDA C., AGÓN C.ASSAYAG G., NTCCRT: A CONCURRENT CONSTRAINT FRAMEWORK FOR SOFT REAL-TIME MUSIC INTERACTION., *Journal of Theoretical & Applied Information Technology*, 82, 1, 2015.

[Tor 16a]   TORO M., Probabilistic Extension to the Concurrent Constraint Factor Oracle Model for Music Improvisation, *ArXiv e-prints*, 2016.

[TOR 16b]   TORO M., DESAINTE-CATHERINE M.CASTET J., An Extension of Interactive Scores for Multimedia Scenarios with Temporal Relations for Micro and Macro Controls, *European Journal of Scientific Research*, 137, 4, 396–409, 2016.

[TOR 16c]    TORO M., PHILIPPOU A., ARBOLEDA S., PUERTA M.VÉLEZ S. C. M., Mean-Field Semantics for a Process Calculus for Spatially-Explicit Ecological Models, MUÑOZ C. A.PÉREZ J. A., , Proceedings of the Eleventh International Workshop on *Developments in Computational Models,* Cali, Colombia, October 28, 2015, 204 *Electronic Proceedings in Theoretical Computer Science*, Open Publishing Association, 79-94, 2016.

[TOR 16d]    TORO M., RUEDA C., AGÓN C.ASSAYAG G., GELISP: A FRAMEWORK TO REPRESENT MUSICAL CONSTRAINT SATISFACTION PROBLEMS AND SEARCH STRATEGIES, *Journal of Theoretical & Applied Information Technology*, 86, 2, 2016.